# Zinc in a +III oxidation state

**Devleena Samanta[1,2] and Purusottam Jena[2]**

[1]Department of Chemistry, Virginia Commonwealth University, Richmond, VA 23284, USA
[2]Department of Physics, Virginia Commonwealth University, Richmond, VA 23284, USA

**Abstract:**

The possibility of Group 12 elements, such as Zn, Cd, and Hg existing in an oxidation state of +III or higher and hence transforming them into transition metals has fascinated chemists for decades. It took nearly 20 years before experiment could confirm the theoretical prediction that Hg indeed can exist in an oxidation state of +IV. While this unusual property of Hg is attributed to the relativistic effects, Zn being much lighter than Hg has not been expected to have an oxidation state higher than +II. Using density functional theory we show that an oxidation state of +III for Zn can be realized by choosing specific ligands with large electron affinities. We demonstrate this by a systematic study of the interaction of Zn with F, $BO_2$, and $AuF_6$ ligands whose electron affinities are progressively higher, namely, 3.4 eV, 4.5 eV, and 8.6 eV, respectively. Discovery of higher oxidation states of elements can help in the formulation of new reactions and hence in the development of new chemistry.

The oxidation state is a "measure of the degree of oxidation of an atom in a substance"[1] and it is the fundamental key to understanding redox reactions, reaction mechanisms, catalysis etc. Transition metals, owing to their incomplete *d*-shells, exhibit variable oxidation states and hence form a large domain of complexes[2]. The possibility of transforming Group 12 elements such as Zn into transition metals has fascinated chemists for decades. Due to Zn's $3d^{10}4s^2$ ground state configuration and highly stable filled *d*-orbitals its inner *d*-electrons seldom take part in bonding and oxidation states of Zn beyond +II are difficult to achieve. As a consequence, relatively fewer compounds of zinc are known. Needless to say, the discovery of new oxidation states of Zn will enable us to formulate new reactions and develop new chemistry. This is particularly important as zinc has many applications in the pure metallic state (in alloys)[3], as salts (used as white pigments)[4], as bio-complexes (metallo-enzymes)[5] and as organometallic reagents (used in organic synthesis)[6].

We realize that the major challenge in achieving an oxidation state of +III and higher for Group 12 elements is to involve their inner *d*-orbitals. This is particularly difficult to accomplish for Zn since its third ionization potential is the largest amongst its congeners[7]. Since this decreases as we go down the periodic table, significant effort has been made in the past to achieve higher oxidation states for the heavier element mercury. In 1976, a short-lived $[Hg^{III}(cyclam)]^{3+}$ species generated through electrochemical oxidation was reported[8].

The +IV state of Hg is expected to be more stable than the +III state since it has the same electronic configuration ($5d^8$) as that of the very stable $Au^{3+}$ cation[9]. Consequently, $HgF_4$, where Hg is in a +IV oxidation state, was theoretically predicted[9-11] about twenty years ago. However, experimental observation eluded scientists until very recently when it was prepared by matrix isolation method[12]. Also, Kaupp et al. have studied weakly coordinating anions as ligands that can stabilize +IV oxidation state of Hg[13]. Unlike $HgF_4$, in all cases they found that $Hg^{IV}$ complexes have at least one exothermic fragmentation pathway. Zn has not yet been shown to exist in an oxidation state of +III or higher.

We wondered if highly oxidizing ligands may enable Zn to exhibit +III oxidation state. Armed with the knowledge that a class of molecules called superhalogens[14] can have electron affinities (EA) far exceeding the value of halogen atoms, we embarked on a systematic study of the interaction of zinc with a variety of atoms and molecules with progressively increasing electron affinities. The high EA of these ligands can be expected to compensate for the large third ionization potential of Zn. We performed a methodical study of the equilibrium geometries and total energies of neutral and anionic $ZnX_3$ clusters for X = F, $BO_2$ and $AuF_6$ using density functional theory. We note that the electron affinities of F, $BO_2$, and $AuF_6$ are, respectively, 3.4[15], 4.5[16], and 8.4[17] eV.

The optimized structures of neutral and anionic $ZnX_3$ are given in Figure 1. $ZnX_3^-$ molecules are expected to be very stable negative ions with large vertical detachment energies (VDE). Table 1 shows the adiabatic detachment energies (ADE) and vertical detachment energies of $ZnX_3^-$ clusters. Indeed, $ZnF_3$ is a superhalogen, $KZnF_3$ being a well-known perovskite salt[18,19]. Similarly, we find $Zn(BO_2)_3$ and $Zn(AuF_6)_3$ to be hyperhalogens[20]. The oxidation state of Zn is +II in these anions. To achieve +III oxidation state, $ZnX_3$ molecules must also be stable as neutrals. The structure, bonding and stability of neutral $ZnX_3$ are discussed below.

**Table 1:** Adiabatic Detachment Energies (ADE) and Vertical Detachment Energies (VDE) of $ZnX_3$ clusters for X=F, $BO_2$ and $AuF_6$

| Cluster | ADE | VDE |
| --- | --- | --- |
| $ZnF_3^-$ | 6.20 | 6.59 |
| $Zn(BO_2)_3^-$ | 5.63 | 6.78 |
| $Zn(AuF_6)_3^-$ | 9.38 | 9.82 |

$ZnF_3$ is planar with $C_{2v}$ symmetry and has two Zn-F bond lengths (1.93 Å and 1.76 Å) signifying two different bond strengths. The F-F bond distance between the two nearest F atoms is 2.04 Å. These results are in good agreement with earlier work on $ZnF_3$ performed at the B3LYP and CCSD(T) levels using effective core potentials, ECP of the Stuttgart group for Zn and aug-cc-pVTZ basis set for F[21]. It is worthwhile to point out that this distance is only slightly longer than the F-F bond distance in $F_2^-$ molecule (2.01 Å). This suggests that the two close F atoms in $ZnF_3$ are quasi-molecular.

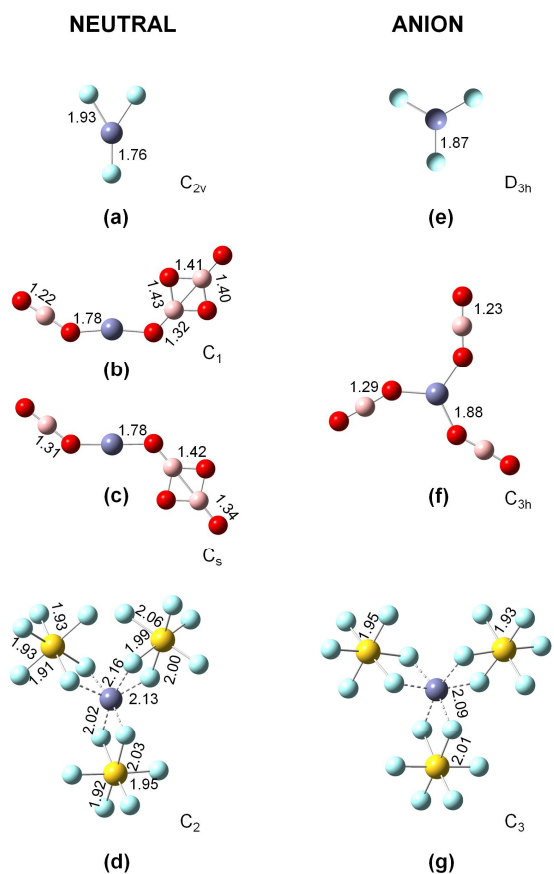

**Figure 1:** Optimized structures of neutral (a-d) and anionic (e-g) ZnX$_3$ (X=F, BO$_2$ and AuF$_6$) clusters

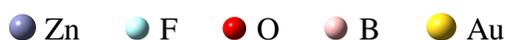

The natural population analysis indicates that the *d*-orbitals of Zn are completely filled in this molecule. The natural bond orbitals (NBO) charge on Zn is 1.682. The charge on one of the F atoms is -0.821 while the charge on each of the other two F atoms which form a quasi-molecule is about -0.430. This indicates that while one 4*s* electron of Zn is transferred to one F atom, the other electron is shared between the remaining two F atoms. The quasi-molecular nature of two of the F atoms in ZnF$_3$ is further demonstrated in Fig. 2 where we compare the charge density contours around the two closest F atoms with that around the F atoms in F$_2^-$. Detailed atom-by-atom NBO charges for all the ZnX$_3$ clusters for X=F, BO$_2$ and AuF$_6$ are listed in Table S1 (see supporting information). It is important to note that the fragmentation of ZnF$_3$ into ZnF$_2$ and ½ F$_2$ is favorable. The reaction is slightly exothermic (by 0.06 eV) which confirms earlier work[21] that it is not a stable compound of Zn in +III oxidation state.

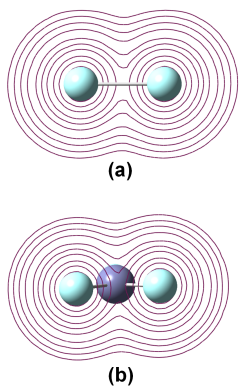

**Figure 2:** Contour diagrams* of (a) isolated $F_2^-$ and (b) two closest F atoms in $ZnF_3$.

*The contour diagrams were plotted using default isovalues in Gaussview 5.0. Identical isovalues were used for both molecules.

Unlike $ZnF_3$, the ground state of $Zn(BO_2)_3$ is stable with respect to fragmentation (see Table S3 in supporting information). However, here Zn is attached to one $BO_2$ ligand and one $B_2O_4$ ligand. This is because two molecules of $BO_2$ dimerize exothermically (-1.70 eV) to form $B_2O_4$. The $BO_2$ and $B_2O_4$ ligands can be both trans and cis to each other giving rise to two degenerate structures. The significance of this result is that Zn is not in +III oxidation state even though there are three $BO_2$ units in the molecule. In both the cis and trans isomers, the charge on Zn is 1.655 while those on the $BO_2$ and $B_2O_4$ units are -0.844 and -0.811 respectively, indicating that the two *s* electrons of Zn are transferred to the two ligands. Furthermore, it is revealed from the natural population analysis that the *d*-orbitals in Zn are completely occupied. We were able to determine a local minimum geometry where all the $BO_2$ moieties are attached individually to Zn i.e. a structure where Zn can be in +III oxidation state. However, this structure is 1.07 eV higher in energy than the ground state. Some energetically low-lying structures of neutral and anionic $Zn(BO_2)_3$ are given in Figure S1 and S2 (see supporting information).

Zn(AuF$_6$)$_3$ has C$_2$ symmetry. Zn is surrounded by three AuF$_6$ units. The three Au atoms form the vertices of an isosceles triangle with the Au-Au distance being 5.92 Å for the equal sides of the triangle and the remaining Au-Au distance being 4.56 Å. The Zn atom is at a distance of 3.15 Å from one Au atom and 3.26 Å from the other two Au atoms. Zn is hexa-coordinated with six F atoms. Also, Zn(AuF$_6$)$_3$ is stable against all the fragmentation channels studied (see Table S4 in supporting information). This indicates that Zn(AuF$_6$)$_3$ is a *stable* molecule of Zn in the +III oxidation state.

Now the question arises if this implies that the inner 3*d* electrons are involved in bonding or not. From the NBO charge distribution it can be seen that one AuF$_6$ moiety has a charge of -0.893 whereas the two AuF$_6$ groups closer to each other have a charge of -0.428 indicating that one *s* electron of Zn is transferred to one AuF$_6$ ligand while the other *s* electron is shared by the remaining two AuF$_6$ groups. This is similar to the case of ZnF$_3$. Again, the natural population analysis suggests that the *d*-orbitals of Zn are completely occupied in Zn(AuF$_6$)$_3$. Why should then Zn be in +III oxidation state in Zn(AuF$_6$)$_3$ while it is in +II oxidation state in ZnF$_3$? It is, therefore, essential to point out a major difference between ZnF$_3$ and Zn(AuF$_6$)$_3$. As stated before, the two closest F atoms in ZnF$_3$ appear to form a quasi-molecular F$_2^-$ unit. Hence, even if ZnF$_3$ were stable with respect to fragmentation, the oxidation state of Zn would still be +II. This possibility is negated in case of Zn(AuF$_6$)$_3$. Though two of the AuF$_6$ moieties are close to each other, they do *not* form a quasi-molecular (AuF$_6$)$_2^-$ unit. This is because (AuF$_6$)$_2^-$ itself is *not* a stable molecule and it fragments into Au$_2$F$_{11}^-$ and ½ F$_2$ releasing 0.42 eV of energy. Similarly, neutral (AuF$_6$)$_2$ is also unstable by 0.80 eV (fragments are 2AuF$_4$ and 2F$_2$). Therefore, Zn(AuF$_6$)$_3$ is a genuine and stable compound of Zn in +III oxidation state. As far as the direct participation of *d*-orbitals in bonding is concerned, we note that NBO charge of +3 in Zn is only possible if bonding is purely ionic. In neutral Zn(AuF$_6$)$_3$ the bonding is not purely ionic, rather a combination of covalent and ionic bonding is featured. In this connection, we note that in AuF$_6^-$, Au exists in a +V oxidation state, but the NBO charge on Au is not +5, but +1.750. Our results imply that to attain the +III oxidation state of Zn, two criteria should be simultaneously satisfied: we require ligands with higher electron affinity and ligands that do not preferentially dimerize. The success of AuF$_6$ as a ligand is attributed to the fulfillment of these two requirements.

It should be mentioned here that Riedel and Kaupp have tried to utilize 'weakly coordinating anions' such as $AlF_4^-$, $Al_2F_7^-$, $AsF_6^-$ etc. to stabilize the +IV oxidation state of Hg[13]. In their work they have also stated that aggregation of the ligands is a major obstacle to achieving higher oxidation states. Furthermore, it is interesting to note that the weakly coordinating ligands tested were all superhalogens, that is, they all have high electron affinities. In fact, we believe that most weakly coordinating ligands are superhalogens and hence they derive their property to stabilize high oxidation states of metals. Our results open the door for the synthesis of new compounds containing metals in unusual oxidation states with potential for applications. It has already been demonstrated that unusually high oxidations states of elements have important consequences. For example, high valent iron (Fe) such as $Fe^{IV}$ (in oxoferryl porphyrins) and $Fe^V$ (in nitridoirons) are important in biochemistry[22] whereas $Fe^{VI}$ (in $FeO_4^{2-}$) has been used for waste water management[23].

According to our calculations, $Zn(AuF_6)_3$ is predicted to be stable at least in the gas phase. It will be a significant milestone in chemistry if it can also be prepared in the condensed phase. $Zn(AuF_6)_2$ is a known salt that has been prepared by the reaction between ultraviolet irradiated $F_2$ and $ZnF_2/2AuF_3$ mixtures in anhydrous HF as solvent. $KrF_2$ can also be used instead of $F_2$. The synthesis has been carried out at room temperature. The product has been characterized using vibrational spectroscopy and X-Ray powder diffraction analysis[24]. The reaction sequence used can be simplified as

$$ZnF_2 + 2\,AuF_3 + 2\,F_2 \xrightarrow[\text{UV}]{\text{aHF}} Zn(AuF_6)_2.$$

We calculated the energies related to the reactants and products in the above reaction. This reaction is exothermic by 5.40 eV. It is well known in chemistry that altering the proportions of the reactants used may give rise to new products. Hence, we suggest that the same reaction may be used to generate $Zn(AuF_6)_3$ simply by altering the ratios of the reactants, such as,

$$ZnF_2 + 3\,AuF_3 + {}^7/_2\,F_2 \xrightarrow[\text{UV}]{\text{aHF}} Zn(AuF_6)_3.$$

This reaction is thermodynamically possible since it is found to be exothermic by 7.20 eV. Of course, for the feasibility of synthesis, entropy is an important parameter which also needs to be taken into account. This obstacle can be overcome by attempting the synthesis at very low temperatures. To aid in characterization, we provide the theoretically predicted IR spectrum for the $Zn(AuF_6)_3$ molecule in Figure S4 in supporting information.

In summary, we have shown that higher and unusual oxidation states of metals can be achieved using ligands with large electron affinities such as superhalogens. We have demonstrated this with the particular case of $Zn(AuF_6)_3$ in which zinc is in a hitherto unknown +III oxidation state. In addition to bearing large electron affinities, it is also important that these ligands have no tendency to dimerize since the contrary would favor fragmentation of the metal-ligand complex. We have also suggested a route through which $Zn(AuF_6)_3$ can be prepared in the condensed phase. Equally important, we show that $ZnF_3$ is a superhalogen with a vertical detachment energy of 6.59 eV while $Zn(AuF_6)_3$ is a hyperhalogen with a vertical detachment energy of 9.82 eV. Consequently, these molecules are predicted to form very stable negative ions. However, in the neutral form, whereas $ZnF_3$ is *not* stable with respect to fragmentation, $Zn(AuF_6)_3$ *is* by 0.26 eV. Though the oxidation state of zinc in the latter molecule is +III, it seems from the NBO charge distribution and natural population analysis that the *d*-orbitals of zinc are not directly involved in bonding. We note that the NBO charge alone can be used to determine the oxidation state *if* the bonding is purely ionic, i.e. there is complete charge transfer between the metal and the ligand. The situation is less clear when the bonding is partly ionic and partly covalent. Nonetheless, our results show that even in the absence of significant direct involvement of *d*-electrons, it is still possible to increase the degree of oxidation of a species and hence form new compounds by using specific ligands. This finding not only demonstrates a way to enhance the chemistry of zinc but also opens the door for the synthesis of unusual compounds by using the strong oxidizing property of superhalogens.

**Theoretical Methods**

All calculations were performed using the Gaussian 03[25] and Gaussian 09[26] packages. The B3LYP[27,28] hybrid functional was used for exchange-correlation potential. The 6-311+G*[29,30] basis set for Zn, F, B and O and Stuttgart pseudopotential SDD[31,32] for Au were used. All geometries were optimized without any symmetry constraints and no imaginary frequencies were obtained. The adiabatic detachment energy (ADE) was determined by calculating the energy difference between the anion ground state of a cluster and its neutral. The vertical detachment energy (VDE), on the other hand, was calculated by taking the energy difference between the anion and the neutral, both at the anion ground state geometry[33]. To understand the nature of bonding involved, Natural Bond Orbitals (NBO) charge distributions were computed. To study the thermodynamic stability of these molecules, the energies associated with different fragmentation pathways were calculated. These are presented in the supporting information (Table S2-S4). Zero-point corrected energies for the lowest energy fragmentation pathways are given in Table S5.

## Acknowledgements

This work is partly funded by a grant from the Department of Energy. We thank NSERC for computational resources.